\documentclass[12pt,preprint]{aastex}

\begin{document} 

\title{The Importance of Dry and Mixed Mergers for Early-Type Galaxy Formation}
\author{S. Khochfar and A. Burkert}
\affil{Max-Planck-Institut f\"ur Astronomie, K\"onigstuhl 17, \\D-69117 Heidelberg, \\ Germany}
\email{khochfar@mpia-hd.mpg.de}
\email{burkert@mpia-hd.mpg.de}
\begin{abstract}
We use semi-analytical modelling techniques to investigate the progenitor
morphologies of present day ellipticals. We find that, independent of the
environment, the fraction of mergers of bulge dominated galaxies
(early-types) increases with time. 
The last major merger of  bright present day  ellipticals
with $M_{B} \lesssim -21$ is preferentially between bulge
dominated galaxies, while those with $ M_{B} \sim -20$ have mainly experienced
last major mergers between a bulge dominated and a disk dominated galaxy. 
 Independent of specific model assumptions, more than $50 \%$
of present day elliptical in clusters with $M_B \lesssim -18$ had last major
mergers which are not of spirals as usually expected within the standard
merger scenario.
\end{abstract} 
\section{INTRODUCTION}\label{int}
The formation of elliptical galaxies by merging disk galaxies has been studied
in numerous simulations  since it was proposed
by \citet{too72} (see \citet{ba92} and \citet{bn03} for reviews). This
merging hypothesis has proven very successful in explaining many of the
properties of 
ellipticals. Even though there are still questions which need further
investigation, like the origin of peculiar core properties of ellipticals, it
is now widely believed 
that ellipticals formed by mergers of disk galaxies. In the framework of
hierarchical structure 
formation,  merging is the natural way in which structure grows. Indeed, the
observed merger fraction of galaxies is in agreement with the predictions of
hierarchical models of galaxy formation \citep{kb01}. Semi-analytical models of
galaxy 
formation, which successfully reproduce many observed properties of galaxies,
generally assume that star formation 
takes place in a galactic disk which formed by gas infall into dark matter
halos \citep[e.g.][]{wf91,kcdw99,sp99,swtk01}. Once these disk 
galaxies merge, depending on the mass ratio of the galaxies, elliptical
galaxies form. N-body simulations suggest a mass ratio of $M_{1}/M_{2}
\le 3.5$, with $M_{1} \ge M_{2}$ to generate ellipticals \citep{nb01}. We
refer to  
these events as major mergers and to events with $M_{1}/M_{2} > 3.5$ as minor
mergers. 
Ellipticals can later on build up new disks by accretion of gas and become
bulges of spiral galaxies \citep[e.g.][]{sn02} or merge with other galaxies.  Up to now the 
frequency of  elliptical-elliptical mergers (dry mergers, e-e) or
spiral-elliptical mergers (mixed mergers, sp-e) has not been studied in detail
despite observational evidence indicating their importance. \citet  {vd99}, for
example, find mergers of red, bulge dominated galaxies in a rich cluster at
intermediate redshifts.  

In this letter we investigate the liklihood of dry and mixed mergers.
 Our semi-analytical
model was constructed similar to those described in detail by \citet{kcdw99}
 and \citet{swtk01}. Merger 
trees of  dark matter halos with different final masses $M_{0}$ at $z=0$ were generated using the method described by \citet{somer99},  which is based on the
 extended Press-Schechter formalism \citep{bond91,bow91}.
 The mass $M_{0}$ traces different environments. We adopt $M_0 = 10^{12}
 M_\odot$ which represents a field environment and $M_0 = 10^{15} M_\odot$
 which is a galactic  cluster environment. 
Present day ellipticals  are identified by their B-band bulge-to-disc ratio as
 in \citet{swtk01}, which corresponds to roughly more than $60 \%$ of the
 stellar mass in the bulge.  We divide the progenitor morphologies into two
 distinct classes. Those with dominant bulge component are labeled {\it e} and
 those with dominant disk component are labeled {\it sp}.  In what follows our standard  model assumes
that the stars of accreted satellites in minor mergers  
contribute to the bulge component of the more massive progenitor and  bulge
dominated galaxies have more than $60 \% $ of their stellar mass in the bulge.
 We adopt a $\Lambda$CDM cosmology with $\Omega_{m}=0.3$,
 $\Omega_{\Lambda}=0.7$ and $H_0 = 65$ km s$^{-1}$ Mpc$^{-1}$. 
\section{MORPHOLOGY OF PROGENITORS}\label{morph}
\begin{figure}
\plotone{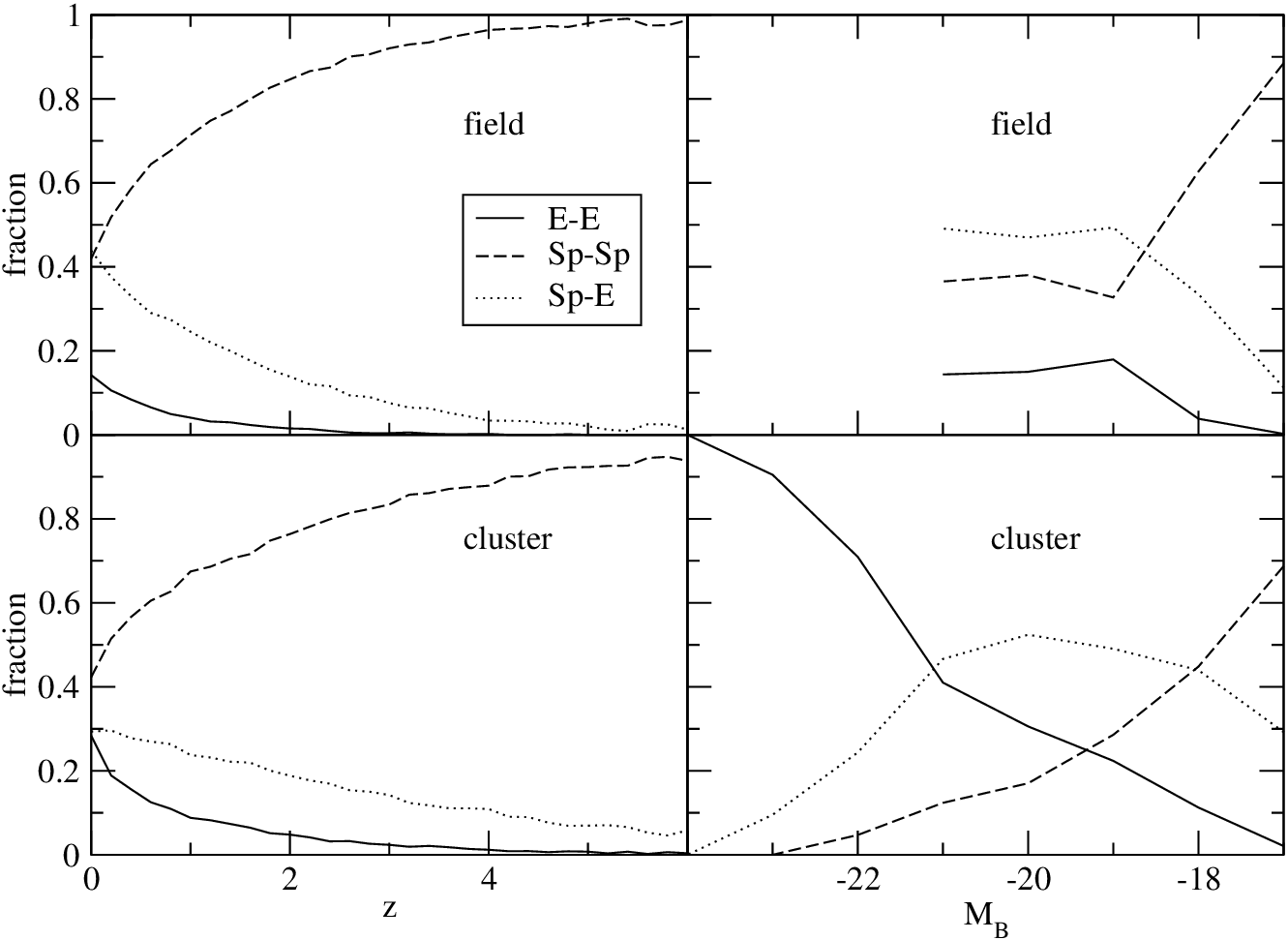}
\caption{Left panel, fraction of major mergers in the standard model between
  galaxies of different 
  morphology at each redshift. Right panel, the fraction of present day
  ellipticals which experienced a last 
  major merger of type sp-sp, e-e or sp-e as function of their B-band
  magnitude.  Results shown for  the standard model.  }\label{pla} 
\end{figure}

We start by analyzing the morphology of  progenitors involved in
major mergers adopting our standard model.
Due to continuous interactions, the fraction of
bulge dominated galaxies increases with decreasing redshift.
As a result, the probability for them to be involved in a major merging event
increases too, which is shown in the left panel of fig. \ref{pla} for a field
($M_0 = 10^{12} M_\odot$) and cluster environment ($M_0 = 10^{15} M_\odot$).
Due to more frequent interactions the increase of the e-e and sp-sp fraction
is faster in more dense environments and
at redshifts $ z \lesssim 1$ the sp-e and e-e fraction show clear environmental
dependencies. The fraction of e-e mergers increases faster (slower) while the
fraction of sp-e mergers increases slower (faster) with time in high density
(low) regions. The most massive
galaxies are mainly bulge dominated \citep [e.g.][]{bm98,k01},  
suggesting that the fraction of e-e and sp-e is mass dependent. The right
panel of fig. \ref{pla}
illustrates the fraction of present day ellipticals at each magnitude which
experienced last major 
mergers of type e-e, sp-sp or sp-e. The fraction of e-e and sp-e mergers indeed
increases towards brighter luminosities with a tendency to increase faster in
more dense environments, due to  the higher fraction of bulge dominated
galaxies. One can distinguish between three luminosity regions: for $M_{B}
\lesssim -21$  dry, at around $ M_{B} \sim -20$,  mixed and for
$M_{B} \gtrsim -18$  sp-sp mergers dominate.    

\begin{figure}
\plotone{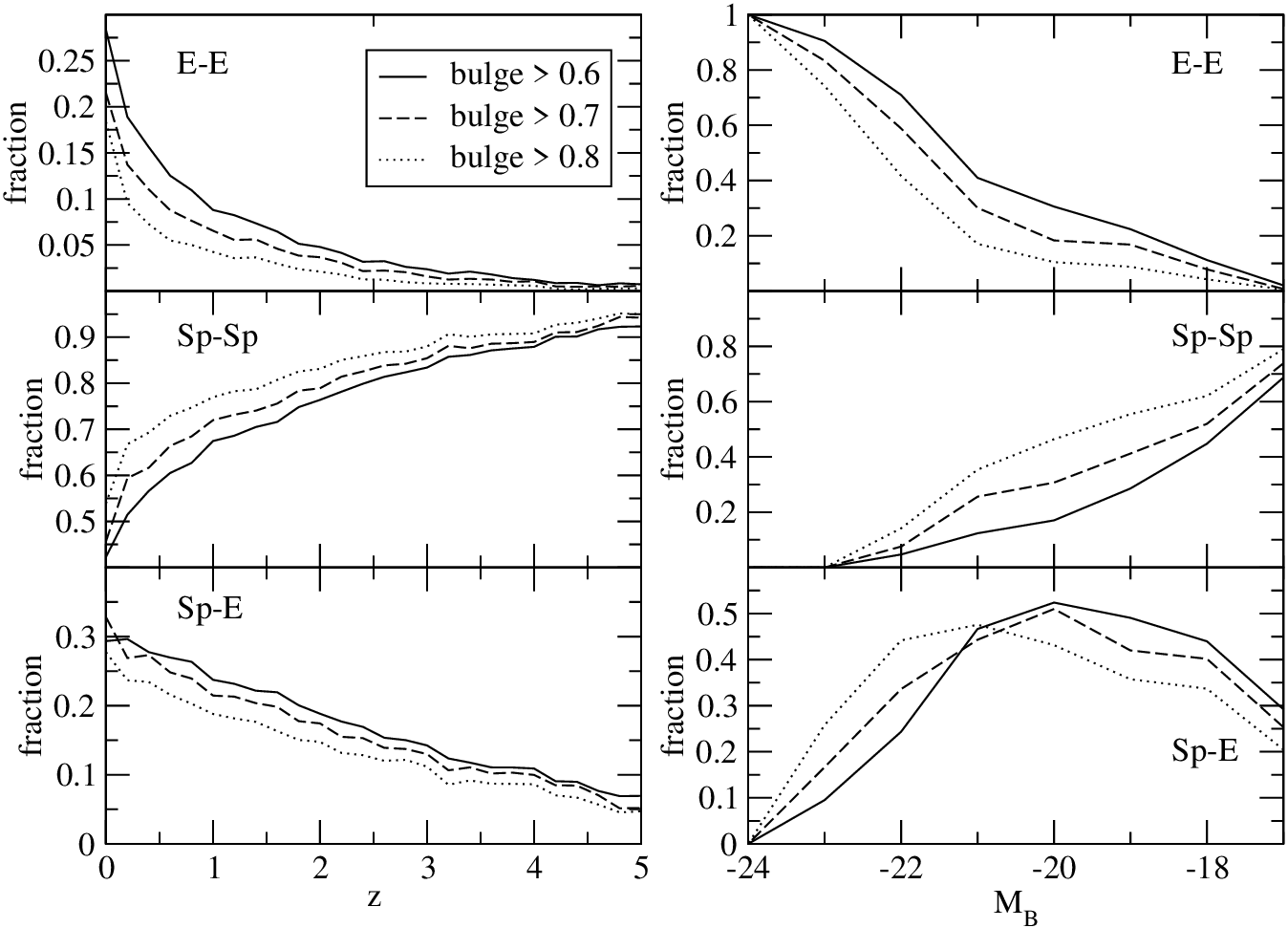}
\caption{The left column shows the dependence of merger fractions of different
  types on the definition of bulge dominated galaxies. The right column displays
  the same dependence for the last major merger type of present day
  ellipticals at each B-band magnitude. Results are shown for a cluster
  environment of $M_{0}=10^{15} M_{\odot}$ and a model where all satellite
  stars from  minor mergers contribute to the bulge of the more massive merger
  partner.  }\label{plc} 
\end{figure}

It is important to understand how  our results depend on the model
assumptions. We focus on cluster environments with
$M_{0}=10^{15} M_{\odot}$, where the fraction of ellipticals is largest, and
investigate the dependence on our definition of 
a bulge dominated galaxy. We varied the definition of a bulge dominated galaxy
from more than $60 \%$ mass in the
bulge component to more than $80 \%$ mass in the bulge. The results are shown
in fig. \ref{plc}. The tighter definition of a bulge dominated galaxy reduces
(increases) 
the fraction of e-e (sp-sp) mergers at all redshifts, which results in a lower
(higher) fraction of last major mergers being between bulge (disk) dominated
galaxies. The right panel of fig. \ref{plc} 
reveals in which mass range the galaxies are most sensitive to the definition
of a  bulge dominated galaxy. At the
high mass end with $M_{B} \lesssim -21$ (e-e region) most of the e-progenitors have
a very large fraction of their mass in their bulge component, while in the
sp-e and sp-sp region the e-progenitors do not have such dominant bulge
components, which explains why the sp-e fraction increases for $M_{B} \lesssim
-21$ if a tighter definition of bulge dominated galaxies is assumed.
  
\begin{figure}
\plotone{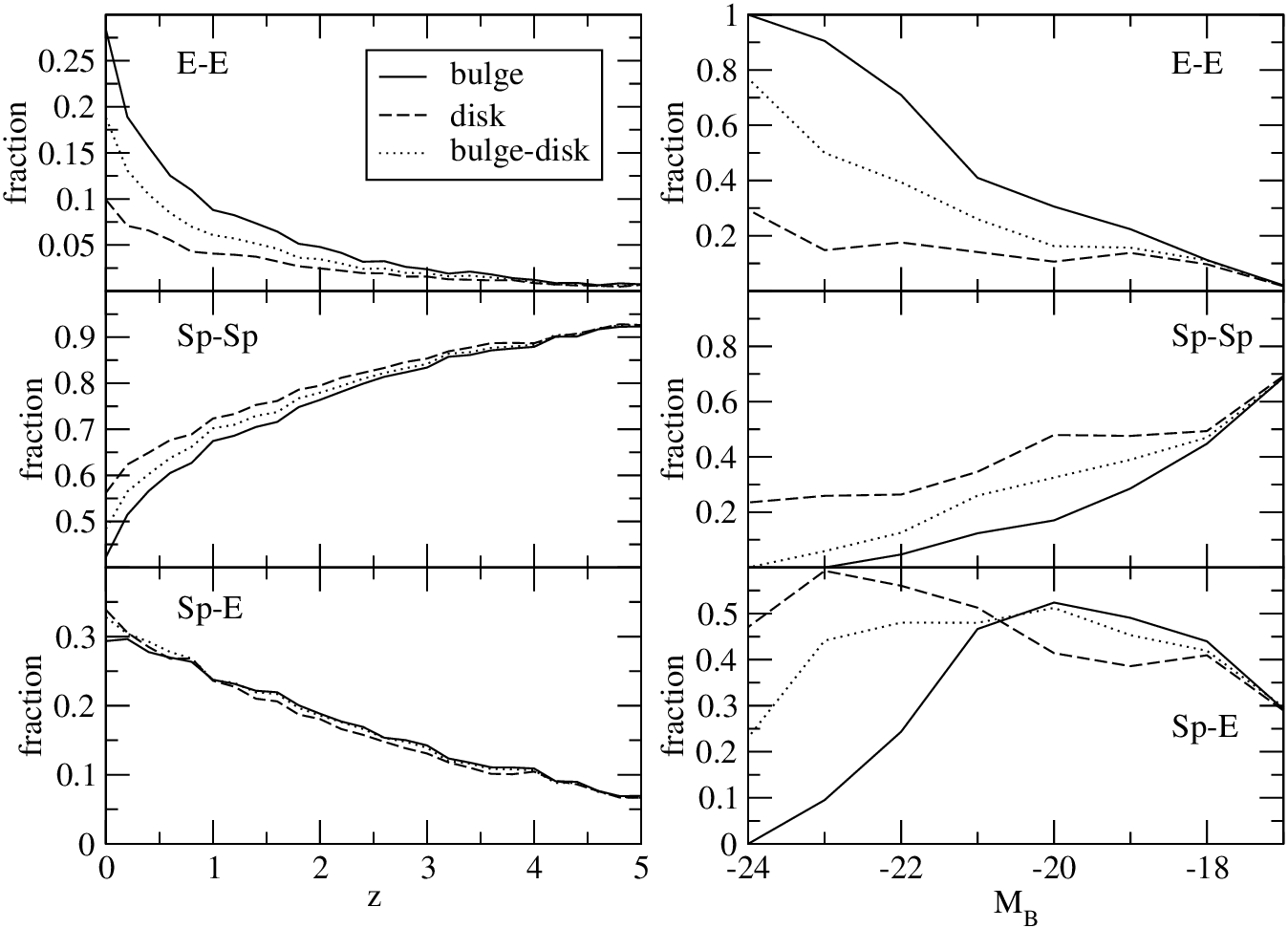}
\caption{The same as fig. \ref{plc}, assuming that galaxies
  with more than $60 \%$ of their mass in the bulge are called ellipticals and
  adopting different fates for 
  the stars of the satellites in minor mergers. We show models where stars
  contribute to the bulge (solid line), to the disk (dashed line) or half of
  the stars to the disk and half to the bulge (dotted line).}\label{pld} 
\end{figure}

 In our standard model we assumed the stars
of a satellite in a minor merger  to contribute to the bulge component of the
more massive progenitor. However the fate of the satellite's stars is not that
clear, as e.g. \citet{wmh96} find that in mergers with $M_1/M_2 =10$ the stars 
of the satellite get added in roughly equal parts to the disk and the
bulge. We tested three different models assuming
the stars of satellites in minor mergers to contribute to the bulge (bulge model)
\citep [e.g.][]{kcdw99},
the disk (disk model) \citep[e.g.][]{sp99}  or half of the stars
to the bulge and the other half to 
the disk (disk-bulge model) of the more massive progenitor. We find that 
the fraction of sp-e merger does not change significant while the fraction of
sp-sp 
(e-e) mergers increases (decreases) from bulge to  disk model (fig. \ref{pld}).
This demonstrates that minor mergers play an important role between two major
merging events of a galaxy. The stars and the gas contributed from the
satellites will affect the morphology of elliptical galaxies  and make them
look more like lenticular galaxies. 

It is interesting to investigate the fraction of present day ellipticals
brighter than a given magnitude which experienced last major mergers of  e-e,
sp-e or sp-sp type.
If bulge dominated galaxies are defined as those with more than $60 \% $
of their mass 
in the bulge. We find, independent of the fate of the satellite stars, that
more than $50 \%$ 
of the ellipticals brighter than  $M_{B}\sim -18$ have experienced a last major
merger which was not a merger between disk dominated galaxies. 

\section{DISCUSSION AND CONCLUSIONS}
We have analyzed the morphologies of progenitors of present day
ellipticals based on their stellar mass content in bulge and disk, finding
that in contrast to the common assumption of disk 
dominated progenitors, a large fraction of 
ellipticals were formed by the merging of a bulge dominated system with a disk
galaxy or another bulge dominated system. \citet{kh00} find that the
fraction of gas involved in the last major merger of present day ellipticals
decreases with stellar mass. We find the same behavior and show in addition
that the fraction of dry  and mixed mergers increases with luminosity,
suggesting that 
massive ellipticals mainly formed by nearly dissipationless mergers of
ellipticals (dry mergers). Our results combined with those of \citet{milo01}
provide an 
explanation for core properties of  ellipticals as observed 
e.g. by \citet{g96}. Progenitors of massive ellipticals should be bulge
dominated with massive black 
holes and very little gas. Their merging leads naturally to flat cores in the
remnant. In contrast, 
progenitors of low mass ellipticals are gas rich with small bulges and low
mass black
holes, resulting in dissipative mergers and cuspy remnants. With these
assumptions it is 
 possible to reproduce the relation between mass deficit and
black hole mass observed by \citet{milo02} (Khochfar \& Burkert in
preparation). It is also interesting
to note that \citet{gen01} and \citet{tac02}  find that ULIRGS have  effective
radii and  velocity dispersions similar to those of intermediate
mass disky ellipticals with $ -18.5 \ge M_B \ge -20.5$ (sp-e
region). QSOs on the other hand have effective radii and velocity dispersions
which are similar to giant boxy ellipticals (e-e region). This suggests that
ULIRGS 
should be formed in sp-e mergers whereas QSOs formed almost dissipationless
through e-e mergers.  

We find that many bulge dominated progenitors experienced minor
mergers in between two major merger events. 
The morphology of these objects is somewhat ambiguous and may
depend on several parameters like the impact parameter of the infalling
satellites. However, it is clear that these galaxies will rather look like
lenticular galaxies than classical spirals. If lenticulars  make up a
large fraction of progenitors of present day ellipticals with $M_B \lesssim
-21 $, numerical simulations of the formation of giant elliptical
galaxies should start with progenitors which were disturbed by minor mergers
and should not use relaxed spiral galaxies \citep[e.g.][]{bn03}. 

Independent of the fate of satellite stars in minor mergers, more than
$50 \%$ of present day ellipticals brighter than $ M_B \sim -18$ in clusters
had a last major 
merger which was not a merger between two classical spiral galaxies. Despite
all the successes of simulations of merging spirals  in explaining
elliptical galaxies our results indicate that only low mass ellipticals are
represented by such 
simulations. More simulations of sp-e \citep[e.g.][]{nb00} and e-e mergers are required to address
the question of the formation of ellipticals via merging adequately.

\clearpage


\end{document}